\documentstyle[epsfig,preprint,prb,aps]{revtex}

\begin{document}
\title{Electronic Structure, Local Moments and Transport in Fe$_{2}$VAl}
\author{D.J. Singh}
\address{Code 6691, Naval Research Laboratory, Washington, DC 20375}
\author{I.I. Mazin}
\address{Code 6691, Naval Research Laboratory, Washington, DC 20375 and CSI, George\\
Mason University, Fairfax, VA 22030}
\date{\today}
\maketitle

\begin{abstract}
Local spin density approximation calculations are used to elucidate
electronic and magnetic properties of Heusler structure Fe$_{2}$VAl. The
compound is found to be a low carrier density semimetal. The Fermi surface
has small hole pockets derived from a triply degenerate Fe derived state at $%
\Gamma $ compensated by an V derived electron pocket at the X point. The
ideal compound is found to be stable against ferromagnetism. Fe impurities
on V sites, however, behave as local moments. Because of the separation of
the hole and electron pockets the RKKY interaction between such local
moments should be rapidly oscillating on the scale of its decay, leading to
the likelihood of spin-glass behavior for moderate concentrations of Fe on V
sites. These features are discussed in relation to experimental observations
of an unusual insulating state in this compound.
\end{abstract}

\subsection{Introduction}

The physics of metal insulator transitions, although much better understood
now than a few years ago, continues to attract interest, especially in
proximity to unusual magnetic behavior or where odd transport properties are
observed on the metallic or insulating sides of the transition.

Nishino et al. \cite{NishinoPRL,NishinoSM} recently reported a highly
unusual insulating state in Heusler phase (Fe$_{1-x}$V$_{x}$)$_{3}$Al alloys
at $x=1/3$. At this composition, the Fe and V are believed to separate on
the two transition metal sub-lattices, yielding a well ordered compound.
The resistivity decreases smoothly from the lowest
measured temperature of 2 K to over 1200 K, characteristic of an insulator.
However, the dependence is clearly non-exponential and has a finite low
temperature value of approximately 3 m$\Omega \cdot $cm. This unusual
resistivity is accompanied by the presence of a clear Fermi edge in
photoemission spectra and a large finite linear component of the specific
heat, $\gamma=14$ mJ/mol$\cdot $K$^{2}$.
It was noted that plots of specific heat
over temperature, $C/T\,\,vs.\,T^{2}$ show an upturn with decreasing
temperature, reminiscent of heavy fermion systems, perhaps related to
spin-fluctuations.

Compositions with slightly less V (lower $x$) have metallic resistivities at
low temperature,  and order magnetically,
while higher V concentration samples display more normal semiconducting
behavior. The ordering is plainly ferromagnetic at low $x\alt 0.2$
and presumably so up to $x=1/3$.  The Curie
temperature, $T_{C}$ decreases monotonically to zero with increasing $x$
from almost 800 K at $x=0$ reaching room temperature near $x=0.2$. In this
ferromagnetic regime, the resistivity as a function of temperature is
metallic though with generally high low temperature saturation values,
has a maximum
at $T_{C}$ and then crosses over to a decreasing $T$ dependence, somewhat
like the colossal magnetoresistive manganites. The prominence of this
cross-over
feature increases rapidly with decreasing $T_{C}$. There is no evidence for
magnetic ordering at $x=1/3$ or higher. However, based on the temperature of
the resistivity maximum as a function of composition, Fe$_{2}$VAl is
at or very close to the composition where $T_{C}$ reaches zero, again
suggestive of strong spin fluctuations.

(Fe$_{1-x}$V$_{x}$)$_{3}$Al alloys occur in a Heusler DO$_{3}$ structure
throughout the composition and temperature range of interest. This structure
is derived from a simple cubic B2 (CsCl) structure, FeAl, by replacing every
second Al by a transition metal atom in an fcc way. Thus it has a single Al
site and two inquivalent transition metal sites in the four atom unit cell.
The first transition metal site,{\it \ }denoted Fe1, accommodates one atom,
and is coordinated by eight transition metal atoms on the other type of
site. The second, Fe2, accommodates two atoms and is coordinated by 4 Al and
four Fe1 transition metal atoms. It is important to note that the Fe1 site
is quite different from the Fe2 site, both in terms of coordination and
size. Particularly, the Fe1 site is larger and has no Al neighbors. Not
unexpectedly, V, being larger, prefers to occupy Fe1 site \cite
{NishinoSM,OBF} leading to the L2$_{1}$ ordered compound Fe$_{2}$VAl. This is
reflected in the sharp upturn in lattice parameter as a function of
composition at $x=1/3$, where the Fe1 site becomes completely V filled, 
resulting in occupation of the Fe2 site with the larger V species.
Nonetheless, Fe$_{2}$VAl is hardly a narrow line compound, and there could
well be significant Fe occupation of the nominal V, Fe1 sites in as
measured $x=0.33$ samples.

In this paper, we report electronic structure studies of the 
(Fe$_{1-x}$V)$_{3}$Al system focussing on the $x=1/3$ composition,
in order to create a framework
for understanding its properties. These local spin density approximation
(LSDA) calculations show fully ordered Fe$_{2}$VAl to be a semimetal,
with separated electron and hole pockets and a very low carrier density.
This is of some interest in itself, since it is a situation favorable for
exciton formation. Remarkably, the compound is not near a ferromagnetic
instability, and in fact has a low spin susceptibility, related to the low
carrier density. However, we find that Fe atoms on the nominal V, Fe1 site
display strong local moment magnetism. The transport and other anomalous
properties are discussed in terms of the band structure and the interactions
with the dilute Fe1 local moment system.

\subsection{Approach}

All atoms in the fcc DO$_{3}$ and ordered L2$_{1}$ crystal structure occur
on high symmetry sites. Accordingly, the only free crystallographic
parameter is the lattice parameter. The LSDA calculations presented here are
based on the reported \cite{NishinoPRL} experimental lattice parameter of 0.576
nm. The electronic structure provides the underlying basis for our
discussion of transport properties, and accordingly we wished to obtain it as
accurately as possible. The band structures, densities of states
and fixed spin moment calculations were all done using the general potential
linearized augmented planewave method \cite{David}, including local orbital
extensions to relax linearization errors, and well converged basis sets of
over 350 functions for the four atom unit cells. \cite{David1} LAPW spheres
of radius 2.30 a.u. were employed for all sites.

The high site symmetries (the lowest symmetry site is the Fe2, which is
tetrahedral) and bcc-like close packing, suggest that linear muffin tin
orbital atomic sphere approximation (LMTO-ASA) calculations may also be
reasonably reliable. We have performed parallel calculations using the
Stuttgart {\it TB LMTO-4.7} code \cite{OKA75,OKA-TB}
to test this. We find that,
apart from small band shifts, the two codes yield identical results. Thus we
were able to use the LMTO method to study the local changes that occur for
transition metal defects in Fe$_{2}$VAl. This was done using a series of
LMTO supercell calculations. In particular we modeled defects in which Fe
atoms were placed on V sites in various magnetic and structural
configurations, in order to establish their magnetic character and
interactions. We used $spd$ orbitals for all atoms, plus downfolded $f$
states on Fe for the LMTO calculations.

\subsection{Electronic structure of ordered stoichiometric Fe$_{2}$VAl.}

\begin{figure}[tbp]                                       
\centerline{\epsfig{file=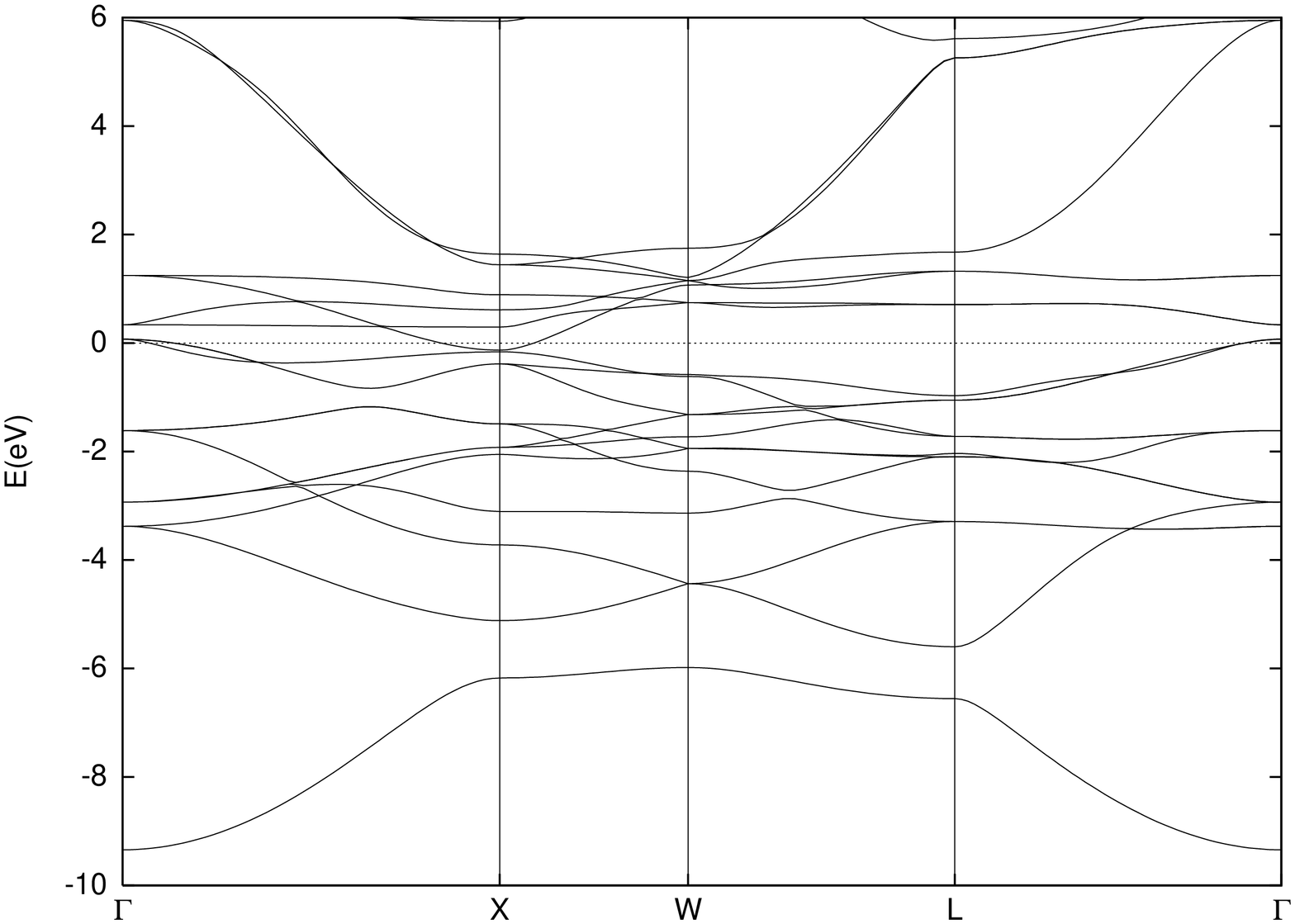,width=0.6\linewidth,angle=-90}}
\vspace{0.1in} \setlength{\columnwidth}{6.5in}                       
\vspace{0.1in}                                                      
\caption{
Calculated band structure of Fe$_2$VAl. The Fermi energy is denoted by
the dashed horizontal line at 0.}
\label{LAPWbands}                                                          
\end{figure}

\begin{figure}[tbp]                                       
\centerline{\epsfig{file=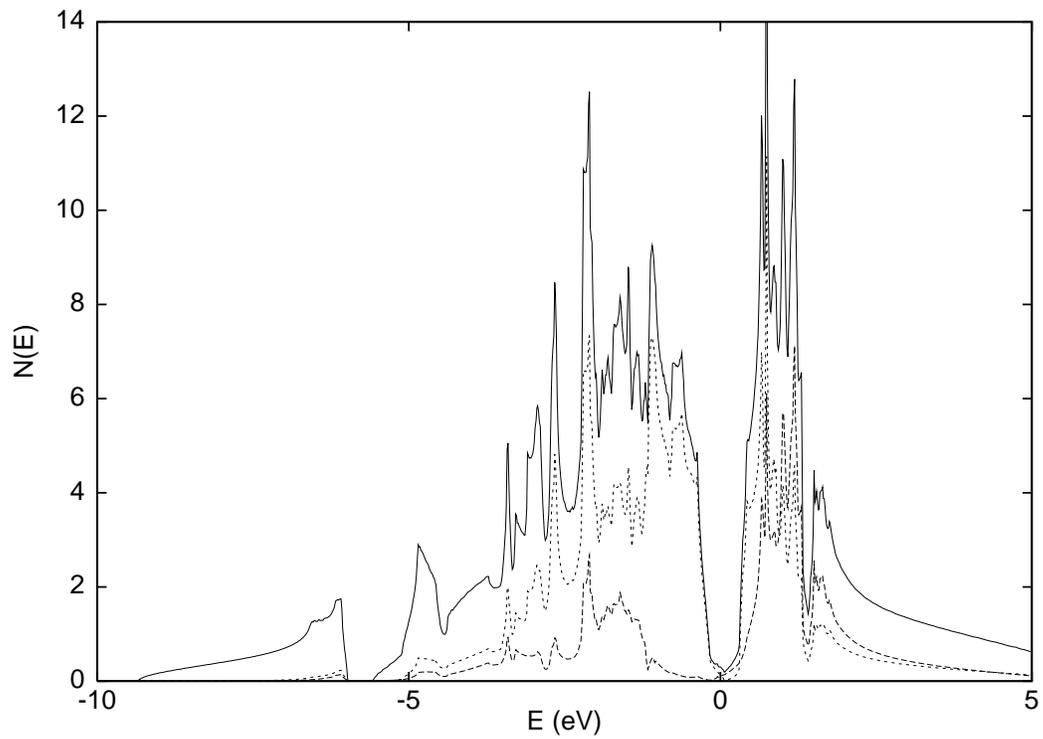,width=0.6\linewidth,angle=-90}}
\vspace{0.1in} \setlength{\columnwidth}{6.5in}                       
\vspace{0.1in}                                                      
\caption{Total and projected DOS of Fe$_2$VAl on a per formula unit basis.
The total DOS is given by the solid line, while Fe d and V d contributions
defined by the projections onto LAPW spheres are given by the dotted and
dashed lines, respectively. The Fermi energy is at 0.}                                   
\label{LAPWDOS}                                                          
\end{figure}

The LSDA band structure of ordered Fe$_{2}$VAl is shown in Fig. \ref
{LAPWbands}. The corresponding electronic density of states (DOS) and
projections onto LAPW spheres is shown in Fig. \ref{LAPWDOS}. Before
discussing the details near the Fermi energy, $E_{F},$ that relate to the
transport properties, we overview the basic structure. No magnetic
instability was found for this composition, and as such the bands are
non-spin-polarized. The 3 eV wide split-off band at the bottom that
disperses upwards from $\Gamma $ derives from the Al 3$s$ state. The
remainder of the valence band manifold, which extends from approximately $-6$
eV to $+2$ eV may be described as 15 transition metal 3$d$ derived bands,
with three Al 3$p$ derived bands dispersing through. However, due to a
Fano-type mixing with the transition metal bands, the Al free electron-like
character is depleted in the main transition metal region and piles up at
the top (2 -- 3 eV, relative to $E_{F}$) and bottom ($-6$ -- $-5$ eV) of the
manifold.

There is a strong pseudogap around $E_{F}$. The states above the
gap are generally of mixed V and Fe $e_{g}$ character, while those below are
for the most-part more Fe-like.

The separation of the V $d$ states into two well-separated peaks in the DOS
is mainly due to crystal field. With the bcc like coordination of the V
site, its $d$ manifold is split into a lower lying set of three $t_{2g}$
states, around $-2$ eV, relative to $E_{F}$, and two $e_{g}$ states around $%
+1$ eV. For the Fe $d$ bands the effects of hybridization are stronger than
the crystal field. This Fe-Fe hybridization involves primarily Al and V
states, although direct hopping is also substantial
 (the Fe - Fe distance is $a/2$:
not much longer than the Fe-V and Fe-Al distances of $a\sqrt{3}/4$). Based
on down-folding of the LMTO band structure, and the positions of the bands
at $\Gamma $, we estimate the Fe crystal field splitting to be quite small,
of the order of 0.35 eV, apparently due to canceling contributions from the
tetrahedral coordinations with V and Al atoms. In any case, as may be seen
from the positions and characters of the bands at $\Gamma $, the ordering of
Fe $d$ sub-levels from lowest to highest
is $e_{g}$ bonding, $t_{2g}$ bonding, $t_{2g}$
anti-bonding and $e_{g}$ anti-bonding. Of these, all but the $e_{g}$
anti-bonding states are below the pseudogap.

However, the pseudogap is not complete because the top of the Fe
anti-bonding $t_{2g}$ band, which occurs at the $\Gamma $ point, is above
the bottom of the V  $e_{g}$ band. The reason is
that V $dd\sigma $ hopping is
comparatively large, and that in an fcc lattice
(the V sublattice in Fe$_{2}$VAl is fcc)
the lower $e_{g}$ band is often strongly
(by 1.5 $t_{dd\sigma })$
dispersive along ${\bf \Gamma }$-{\bf X} direction, so the bottom of
this V-derived band (at the {\bf X} point) occurs below the top of the
anti-bonding Fe $t_{2g}$ band (at the ${\bf \Gamma }$ point)

\begin{figure}[tbp]                                       
\centerline{\epsfig{file=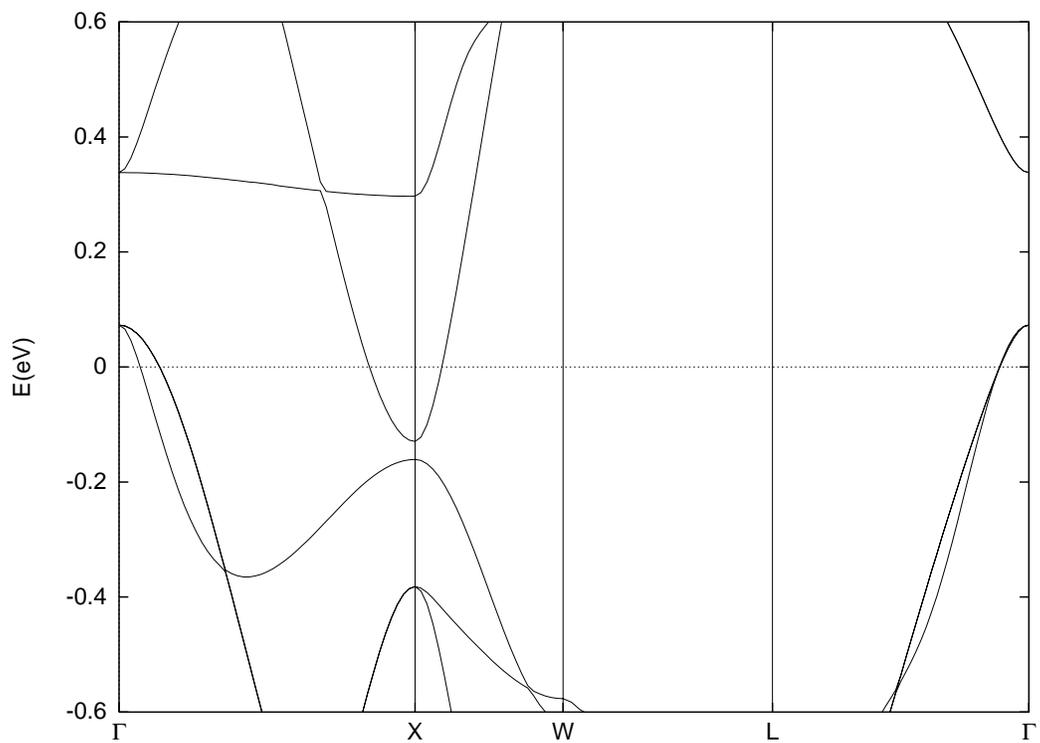,width=0.6\linewidth,angle=-90}}
\vspace{0.1in} \setlength{\columnwidth}{6.5in}                       
\vspace{0.1in}                                                      
\caption{Blowup of the band structure of Fe$_2$VAl near E$_F$.}
\label{blowup}                                                          
\end{figure}

Fig.\ref{blowup} shows a blow-up of the band structure near $E_{F}$. There
are four bands crossing $E_{F}$. These contribute three small hole pocket
sheets of Fermi surface centered at $\Gamma $, compensated by a larger
electron like Fermi surface section centered at the X point of the fcc
Brillouin zone. The occupation is 0.012 electrons/f.u. (2.5$\times 10^{20}$
el./cm$^{3}),$ compensated by an equal number of holes. The $\Gamma $
centered surfaces derive from the Fe $t_{2g}$ bands with a small admixture
of V character. The electron pocket at X is purely V $e_{g}$ (there is no Fe 
$d$ character at the X point by symmetry). The  $\Gamma $ hole
surfaces consist of three pockets with the effective mass $m\approx
0.5$ $m_e$; however, the surfaces deviate noticeably from the sphere, 
due to the finite, albeit small, band filling, and 
the effective masses at the Fermi level vary from 0.35 $m_e$ to 1 $m_e$.
The X centered electron sections are ellipsoids with the
 effective masses  $m_x=0.55$ $m_{e}$ and $m_y=0.25$ $m_{e}$.
The low carrier density corresponds to a low density of states, $%
N(E_{F})=0.3$ eV$^{-1}$ with a bare specific heat coefficient, $\gamma _{%
{\rm bare}}=0.65$ mJ/mol$\cdot $K$^{2}$. Comparing with the experimental
value, this yields a large enhancement of more than 20,
 certainly not characteristic
of a simple metal.

Two candidate mechanisms for yielding such behavior are
strong spin-fluctuations and strong electron-electron correlations. With
regard to the latter, it is noteworthy that the band structure near $E_{F}$
consists of small separated electron and hole sections with low carrier
density, and that there are manifolds of flat transition metal $d$ bands in
close proximity both above and below the pseudogap. The formation of bound
electron-hole pairs (excitons) leading to a correlated insulating state in
zero gap semiconductors and semimetals was much discussed in the 1970's. \cite
{ei} Qualitatively, the formation of such a state in semimetals is possible
when certain conditions are met. These are (a) low enough carrier density
for the pairs not to overlap; (b) exciton binding energy larger than the
distance from the band edge to $E_{F}$ (if scattering is significant, or a
more complicated criterion related to nesting of the hole and electron
sections if the mean free path is long\cite{KM}); and (c) mean free path
long compared to the exciton radius. Given the experimental data,
particularly indications of spin fluctuations (which would scatter carriers),
the high residual resistivities, and smooth behavior as a
function of $x$, it seems unlikely that these
conditions are met in Fe$_{2}$VAl.

The low carrier density implies that a simple Stoner instability of the
non-spin-polarized state, due to divergence of the susceptibility $\chi (q),$
will not occur, since the $N(E_{F})$ factor will be too small. However, the
absence of an instability against small fluctuations does not necessarily
mean that a magnetic state with larger moments is not present. A local
moment system on the verge of ordering is consistent with many of the
experimental observations and would have strong spin fluctuations that could
yield the size of enhancements observed.

\begin{figure}[tbp]                                       
\centerline{\epsfig{file=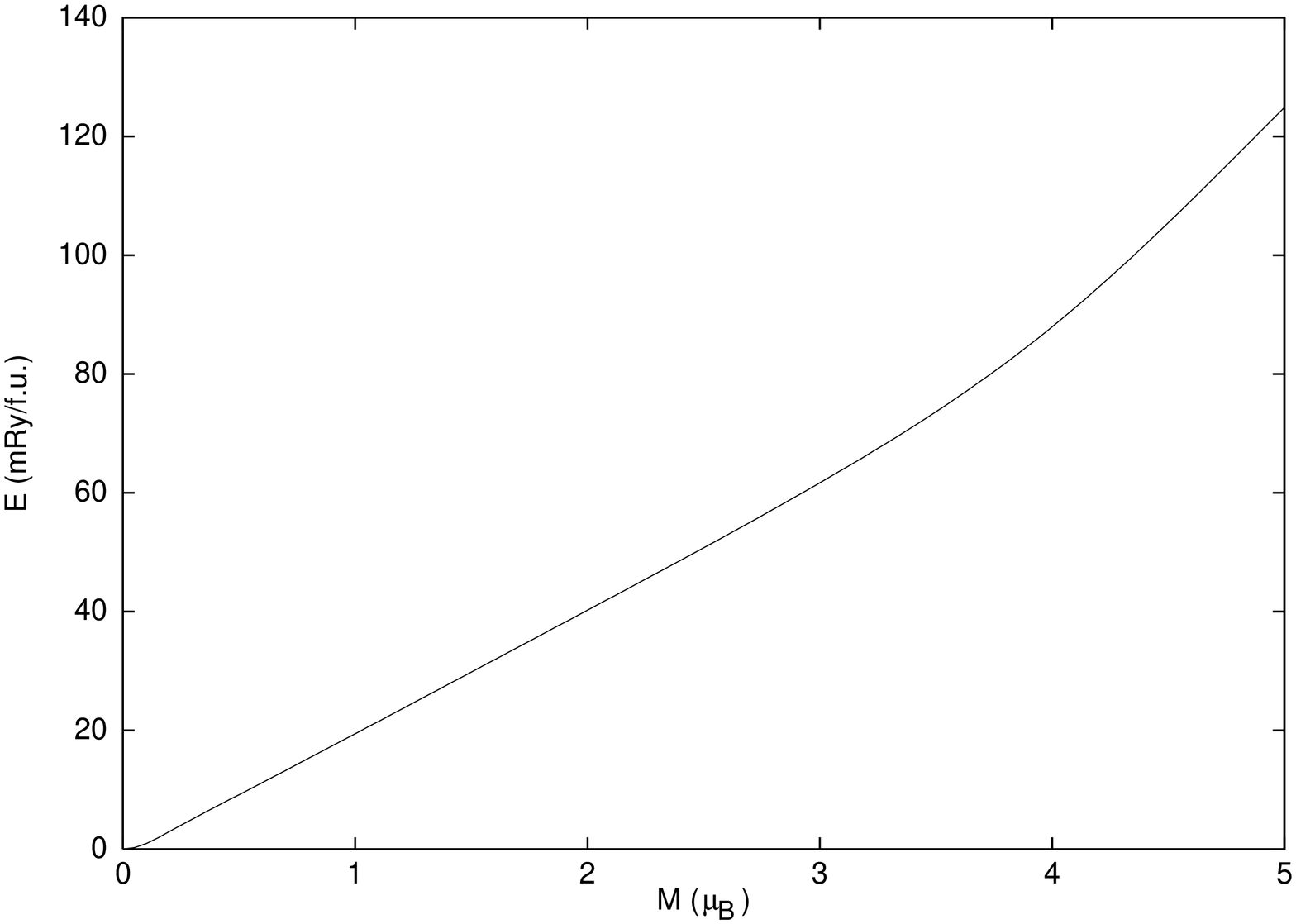,width=0.6\linewidth,angle=-90}}
\vspace{0.1in} \setlength{\columnwidth}{6.5in}                       
\vspace{0.1in}                                                      
\caption{Energy vs. imposed magnetization from fixed spin moment
calculations for stoichiometric Fe$_2$VAl on a per formula unit basis.}
\label{fixedM}                                                          
\end{figure}

Fixed spin moment calculations were used to address this possibility. The
energy as a function of magnetization, shown in Fig.\ref{fixedM}, provides
no evidence of any interesting magnetic behavior in ordered stoichiometric
Fe$_{2}$VAl. In
particular, the energy is a smooth monotonically increasing function of
magnetization. After a very small parabolic region extending to
approximately 0.02 $\mu _{B}$/f.u., the curve becomes roughly linear up to
2.5 $\mu _{B}$/f.u., and then crosses over to an upward curving form. The
linear region is due to the semi-metallic character of the material. As the
bands are split by the exchange enhanced applied field in the fixed spin
moment calculation, $N(E_{F})$ increases from a low value. This leads to a
corresponding increase in the differential susceptibility, flattening the
curve from parabolic. The induced moments are almost entirely associated
with the Fe sub-lattice in Fe$_{2}$VAl. Up to 4 $\mu _{B}/$f.u., the V site
contributes a small ($\sim $5\% of the total) negative contribution, while
by 5 $\mu _{B}$/f.u., the V contribution becomes parallel to the Fe, but is
still very small.

\subsection{Magnetic Properties of Defects}

This begs the question of the magnetic properties of defects, since some
explanation of the magnetic properties is needed. As mentioned, when all the
V on the Fe1 site is replaced by Fe, i.e. in Fe$_{3}$Al, the material is
strongly ferromagnetic. As V is put back in (entering the larger Fe1 site),
ferromagnetism is gradually suppressed, disappearing just at the point where
the Fe1 site is fully substituted with V. On the other hand, the above fixed
spin moment calculations indicate that V is magnetically inactive on the
Fe1 site. One may then conjecture that magnetism is coming from the Fe atoms
on the Fe1 site. Although little is known about defects in Fe$_{2}$VAl, if
it follows the pattern of other compound aluminides, defects in which the
small atom (Fe) sits on the larger sites (Al and V), either accompanied by
non-stoichiometry or compensating vacancies are likely. There are some
experimental indications \cite{Moss} that this is indeed the case in
Fe$_{2+x}$V$_{1-x}$Al.
Given the variation of lattice parameter with composition,
showing a sharp increase at $x=1/3$ and the wide range of V/Fe compositions
that can be made, we conjecture that Fe atoms on the nominally V, Fe1 site
may be a common defect in the experimentally studied samples.

\begin{figure}[tbp]                                       
\centerline{\epsfig{file=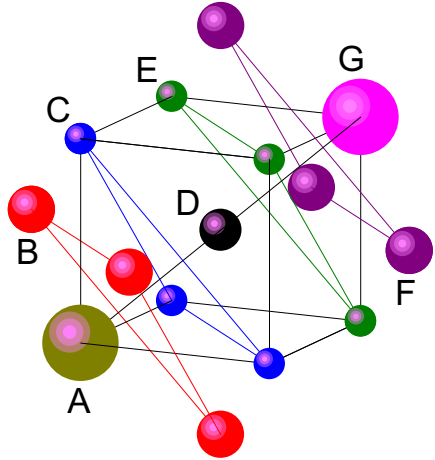,width=0.66\linewidth}}
\vspace{0.1in} \setlength{\columnwidth}{6.5in}                       
\vspace{0.1in}                                                      
\caption{Illustration of the coordination in Heusler type lattices,
labeled as in the supercell calculations (see text).}
\label{STRUCT}
\end{figure}

To study the behavior of such defects, we performed a series of supercell
calculations for various Fe$_{2+x}$V$_{1-x}$Al compounds, using the LMTO-ASA
method. Most of the supercells were produced
by doubling or quadrupling the unit cell along the $(111)$ direction,
resulting in a rhombohedral symmetry. A convenient nomenclature for such
supercells is obtained by listing the sequence of the $[111]$ planes.
For this type of supercell, the crystallography of the Heusler structure is
such that, given a sequence of the $[111]$ planes $\cdots $ABCDEFG$\cdots ,$
the atom D has 3 nearest neighbors (n.n.) of the kind C, 3 n.n. of the kind
E, 1 n.n. of the kind A, and 1 of the kind G (but no n.n. in planes B or F),
as illustrated in Fig. \ref{STRUCT}
Thus, such a notation lets one
immediately see the environment of each atom (recall that every atom sits
in the center of a cube formed by its 8 n.n. as in the B2, CsCl structure).

To start with, let us look at the stoichiometric compound Fe$_{2}$VAl, but
placing V on one of the Fe2 sites, and one of the Fe's on the Fe1 site. This
results in the following sequence:
\begin{equation}
\cdots {\rm V}|{\rm Fe}^{\prime }{\rm AlFe\underline{{\rm FeV}}AlFe}^{\prime
\prime }{\rm V|Fe}^{\prime }\cdots ,  \label{cell1}
\end{equation}
where the underlined atoms have been interchanged. We find a
ferromagnetic ordered state with the moments inside the atomic spheres
2.1 $\mu _{B}$ on \underline{Fe}, 1.1 $\mu _{B}$ on Fe$^{\prime \prime },$
0.6 $\mu _{B}$ on Fe$^{\prime },$ and $-0.2$ $\mu _{B}$ on V (note the minus
sign, in accord with the fixed-moment LAPW calculations). All other atoms
carry negligible moments. This is to be compared with the moments in pure
Fe$_{3}$Al in LMTO-ASA calculations,
which are 2.3 $\mu _{B}$ for Fe1 and 1.8 $\mu _{B}$ for Fe2;
see also Ref.\cite{kud}
(LAPW calculations yield 2.4 and 1.9 $\mu _{B}$, respectively).
Fe on Fe1 sites that are magnetically active and in turn polarize
Fe2 Fe atoms.
Magnetic moments on Fe1, but not on Fe2, are essentially local
moments: they hardly depends on the Fe1's environment or electron count as
is illustrated by the other supercell calculations, described below.
Fe$^{\prime }$
and Fe$^{\prime \prime }$ in (\ref{cell1}) have the same n.n. environment,
but different next n.n. environments. Correspondingly, the induced
magnetization, totaling 1.8 $\mu _{B}$ per 2 atoms, is unevenly distributed
between Fe$^{\prime }$ and Fe$^{\prime \prime }.$ In real samples one expects
that V on Fe2 site, if any, may occupy all Fe2 sites randomly. To investigate
this, we repeated the above calculations with a virtual crystal of
the structure
\begin{equation}
\cdots {\rm V}|{\cal M}^{\prime \prime }{\rm Al{\cal M}}^{\prime }{\rm 
\underline{{\rm Fe}}{\cal M}}^{\prime }{\rm Al{\cal M}}^{\prime \prime }{\rm %
V|{\cal M}}^{\prime \prime }\cdots ,
\end{equation}
where ${\cal M}$ has atomic number $Z=25.25$ (V$_{0.25}$Fe$_{0.75}),$ still
isoelectronic with Fe$_{2}$VAl.

The moment on \underline{Fe} was again 2.1 $\mu _{B},$ that on ${\rm {\cal M}%
}^{\prime \prime }$ was 0.75 $\mu _{B},$ and on ${\rm {\cal M}}^{\prime }$
it was 0.15 $\mu _{B}.$ In other words, although now the moment was equally
distributed between the two ${\rm {\cal M}}^{\prime \prime }$ atoms, the
total magnetization induced in this sublattice changed very little. An
interesting question is why ${\rm {\cal M}}^{\prime }$ atoms, which have
among n.n. three strongly polarized Fe atoms, are not magnetic, while the  $%
{\rm {\cal M}}^{\prime \prime }$ atoms, having only one spin-polarized n.n.,
are.

It is instructive to see how Fe on Fe1 sites behaves in more Fe-rich compounds,
particularly, how defects in which V is placed into a smaller Fe2 site
(as in the
above-considered supercells) influence the magnetic properties?
To answer this question, we prepared supercells with
nominal stoichiometries,
Fe$_{2.5}$V$_{0.5}$Al and Fe$_{2.25}$V$_{0.75}$Al. The
former was, in the above notation,
\begin{equation}
\cdots {\rm V}|{\rm Fe}^{\prime \prime }{\rm AlFe}^{\prime }{\rm \underline{%
{\rm Fe}}Fe}^{\prime }{\rm AlFe}^{\prime \prime }{\rm V|Fe}^{\prime \prime
}\cdots ,  \label{cell2}
\end{equation}
while the latter was a quadrupled fcc cell and retaining full cubic
symmetry. In the former case we obtain $M=2.4$ $\mu _{B}$ on \underline{Fe}
(very slightly larger than in Fe$_{3}$Al) and 0.7 $\mu _{B}$ on Fe$^{\prime
\prime },$ in accord with the virtual crystal calculation described above.
The Fe$_{2.25}$V$_{0.75}$Al compound produced 2.2 $\mu _{B}$ on \underline{Fe%
} with no significant polarization on the other sites.

Taken together, these calculations indicate that Fe on a Fe1 site
acquires strong, localized magnetic moment of 2.2--2.3 $\mu _{B}$, which is
robust against redistribution of Fe and V atoms within the transition metal
sublattices and even changes in the total V concentration. As a further test
we performed supercell calculations with a
further doubling of the unit cell (\ref{cell2}) of Fe$_{2.5}$V$_{0.5}$Al
and performed calculations with antiferromagnetic ordering:
\begin{equation}
\cdots ({\rm Fe}^{\prime \prime }{\rm AlFe}^{\prime }{\rm \underline{{\rm Fe}%
}Fe}^{\prime }{\rm AlFe}^{\prime \prime }{\rm V)}[{\rm Fe}^{\prime \prime }%
{\rm AlFe}^{\prime }{\rm \underline{{\rm Fe}}Fe}^{\prime }{\rm AlFe}^{\prime
\prime }{\rm V]}\cdots .
\end{equation}
We found a metastable antiferromagnetic
self-consistent solution with \underline{Fe%
} having $\pm 2.3$ $\mu _{B}$ and Fe$^{\prime \prime }$ having $\pm 0.6$ $%
\mu _{B}.$ That is, the moment on \underline{Fe} was virtually the same as in
the ferromagnetic case, and that on Fe$^{\prime \prime }$ was only slightly
suppressed. It seems that,
independent of the spin arrangement of other Fe atoms,
Fe on Fe1 site has enough room to act like a quasi-free ion, with the
crystal splitting larger that the $e_{g}$ and $t_{2g}$ subband widths, and
thus is unavoidably magnetic. Fe on the Fe2 site, on the other hand, is
squeezed by larger Al and V ions and has subband widths large compared to
their splitting. Thus it is intrinsically nonmagnetic, although moments can
be induced on this site. In fact, in all supercell calculations, the
partial DOS on \underline{Fe} has two clear maxima, corresponding to the
$e_{g}$ and $t_{2g}$ subbands (as in the non-spin-polarized case), while the
DOS on the other Fe atoms shows relatively structureless broad bands.

\subsection{Discussion}

The above calculations show two main features. First of all Fe$_2$VAl
is semi-metallic with a low carrier density and well separated hole
and electron Fermi surface sections. Secondly, defects in which Fe atoms
occur on the nominally V Fe1 sites provide local moments. Although
this in itself does not provide an explanation of the anomalous 
properties, we speculate
that the odd properties of this system may be due to the dynamics of
a dilute system of such local moments and their interaction with the 
charge carriers.

An interesting point, worth mentioning, is that the direct exchange
interaction of the localized Fe1 moments should be very small. One may
then ask if there is an oscillating RKKY type
interaction that could lead to a spin glass
state in the same manner as in classical spin glass systems. The fact that the
Fermi surfaces are small does not mean that the period of the RKKY
interaction in this system is very long of the order of 2$\pi /k_{F};$
rather, it is
controlled by the wave vector connecting the hole and electron Fermi surfaces,
which is $\pi /a.$ That is, the relevant scale is set not by the size of the
Fermi surfaces, but their separation.
In fact, the RKKY interaction in this system can be written
as 
\begin{equation}
H_{RKKY}({\bf r})\approx 6H_{RKKY}^{0}({\bf r})(\cos x/a+\cos y/a+\cos z/a),
\end{equation}
where $H_{RKKY}^{0}({\bf r})$ is the standard RKKY interaction\cite{Kittel},
a function of $k_{F}$ and effective mass and the factor 3 appears because of
degeneracy of the hole Fermi surfaces. $H_{RKKY}^{0}({\bf r})$
changes with {\bf r} as $\cos (k_{F}r),$ where $k_{F}\approx 0.1\pi /a,$ and
thus is a long wavelength modulation.
The RKKY interaction has a prefactor proportional to $k_{F}^{4};$ one effect of
the smallness of the Fermi surfaces is that the interaction will be fairly
weak compared with the traditional metallic RKKY
spin glasses. Nevertheless, spin-glass effects cannot be
excluded and it is tempting to ascribe the increase of the resistivity at
$T\rightarrow 0$, and anomalously large specific heat coefficient $\gamma $,
to a spin glass transition with a freezing temperature close to zero. It is
worth noting that the temperature dependence $\gamma $ is also unusual, with
a large negative $T^{2}$ term. In this scenario, the properties will be
sensitive to the ordering and concentration of local moments, i.e. Fe on
the Fe1 site. Presumably this will be reflected in a high sensitivity to the
exact composition and growth conditions. In this picture the magnetic
properties of Fe rich alloys near $x=1/3$ are then those of a local moment
system with concentration increasing as $x$ is reduced.

\subsection{Acknowledgments}

Computations were performed using facilities of the DoD HPCMO ASC Center.
Work at the Naval Research Laboratory is
supported by the Office of the Naval Research.

\end{document}